# Tunable, all-optical quasi-monochromatic Thomson X-ray source

K.Khrennikov[1,2], J. Wenz[1,2], A. Buck[2], J. Xu[2,3], M. Heigoldt[1,2], L. Veisz[2] and S. Karsch[1,2]

[1]*Ludwig-Maximilians-Universität München, Am Coulombwall 1, 85748 Garching, Germany*

[2]*MPI für Quantenoptik, Hans-Kopfermann-Str. 1, 85748 Garching, Germany*

[3]*State Key Laboratory for High Field Laser Physics, Shanghai Institute for Optics and Fine Mechanics, Shanghai 201800, China*

**Brilliant X-ray sources are of great interest for many research fields from biology via medicine to material research. The quest for a cost-effective, brilliant source with unprecedented temporal resolution has led to the recent realization of various high-intensity-laser-driven X-ray beam sources[1-7].**

**Here we demonstrate the first all-laser-driven, energy-tunable and quasi-monochromatic X-ray source based on Thomson backscattering. This is a decisive step beyond previous results, where the emitted radiation exhibited an uncontrolled broad energy distribution. In the experiment, one part of the laser beam was used to drive a few-fs bunch of quasi-monoenergetic electrons from a Laser-Wakefield Accelerator (LWFA), while the remainder was scattered off the bunch in a near-counter-propagating geometry. When the electron energy was tuned from 10-50 MeV, narrow-bandwidth X-ray spectra peaking at 5-35keV were directly measured, limited in photon energy by the sensitivity curve of our X-ray detector. Due to the ultrashort LWFA electron bunches, these beams exhibit few-fs pulse duration.**

In recent years, laser wakefield electron acceleration[8-11] has matured enough to produce low emittance[12,13,14] energy-tunable[15,16] MeV-GeV-scale[17,18], few-fs[19,20] electron bunches with good stability. These beams are of great interest as a driver for various free electron X-ray sources[21,22]. Transversely wiggling these beams in alternating or strongly focusing fields produces brilliant X-ray beams as undulator[1,2], betatron[3,4] and Thomson[5-7] radiation.

Although the spectral bandwidth of LWFA currently is still too large for making them useful as a driver for an X-ray FEL, their ultrashort duration in combination with high beam charge and low emittance has yet to be matched by conventional RF accelerators. These qualities promise peak brilliances several orders of magnitude above the current state-of-the art for incoherent, hard X-ray sources.

As an example, the ELBE RF-linac team at Forschungszentrum Dresden-Rossendorf recently reported the generation of 13 (thirteen) 12.3-keVThomson-scattering photons/shot in a 1.13 µsr solid angle from a 22.5 MeV electron beam with 4 ps duration[23]. Our LWFA-driven source produces 100 times more photons/solid angle from a less than 3.5 times smaller spot, in a 1000 times shorter pulse, albeit with an approx. 4-fold energy spread. From these numbers it becomes clear that LWFA beams surpass current RF-accelerated X-ray beams by more than 5 orders of magnitude in brilliance even in this very early state of development.

Of the three X-ray generation schemes mentioned above, Thomson scattering offers the most promising route to well-controlled, keV to MeV X-ray beams due to its short wiggling period. Neglecting higher harmonics, the energy of the emitted photons for the case of near-head-on collision of the laser pulse and a bunch of highly relativistic electrons is given by[24,25]:

$$\hbar\omega_{X-ray} = \hbar\omega_{Laser} \frac{2(1+\beta cos\theta_I)\gamma^2}{1+a_0^2/2+\gamma^2\theta_O^2} \approx 1.55 \text{ eV} \frac{4\gamma^2}{1+a_0^2/2+\gamma^2\theta_O^2} \qquad (1)$$

Here $\omega_{X-ray}, \omega_{Laser}$ are the frequencies of scattered radiation and colliding pulse, $\gamma$ and $\beta$ are the relativistic quantities of the electrons. $a_0 = eE_{Laser}/m_e\omega_{Laser}c$ is the normalized vector potential of the wiggling field, where $e, m_e, E_{Laser}, c$ are the electron charge, rest mass, laser electric field and speed of light, respectively. $\theta_I$ and $\theta_O$ are the interaction and observation angles. The emitted radiation is monochromatic for collimated, monoenergetic electron and laser beams, $a_0 << 1$ and an infinitely small observation area. Since experiments in the past primarily lacked sufficiently monoenergetic laser-driven electron beams, up to date no quasi-monochromatic X-ray beams were observed[1] from an all-optical source.

The number of generated X-ray photons $N_X$ per shot is governed by the Thomson cross-section $\sigma_{th}$, the number of electrons $N_e$ and laser photons $N_{Laser}$. $r_0$ is the beam radius at the interaction region assuming a matched electron and laser beam size:

$$N_X = \frac{1}{\pi r_0^2} \sigma_{Th} N_e N_{Laser}; \qquad \sigma_{Th} = 6.7 \times 10^{-29} m^2 \qquad (2)$$

The tight spot size of LWFA-bunches close to the source is therefore an important asset for achieving high X-ray flux. The duration of the scattered X-ray pulse $\Delta T_X$ is governed by the convolution of electron bunch envelope with a two times Lorentz-contracted colliding laser pulse envelope. For the case both being temporarily nearly Gaussian, one obtains:

$$\Delta T_X \approx \sqrt{(\Delta T_{el})^2 + (\Delta T_{Laser}/4\gamma^2)^2} \qquad (3)$$

Here $\Delta T_{el}$ and $\Delta T_{laser}$ are the FWHM electron bunch and laser pulse durations. Due to the double Lorentz contraction of the laser pulse, for high electron energies the X-ray pulse closely reflects the electron pulse duration of a few femtoseconds[19,20].

In the setup depicted in Fig. 1, the electron bunches for radiating the X-rays were produced by LWFA in an energy-tunable, quasi-monoenergetic fashion (see Fig. 3 left). This can be achieved by controlling the electron injection into the laser-driven wakefield via the shock-front injection scheme[16,26]. The electron energy is easily controlled by moving the injection point along the laser propagation axis inside the plasma (see methods). In the current setup, electron beams with an energy peak tunable between 10 MeV and 150 MeV with a constant energy bandwidth of 5 MeV FWHM are created, which contain an average charge of 20 pC. For a limited energy range the beam divergence was measured during this campaign. It ranged from 17.5 mrad at 27 MeV to 12 mrad at 45 MeV. This is comparable with the values reported for a wider parameter range in Ref. 16, where the electron production was investigated under the same conditions as here.

The colliding pulse is focused to a spot size of 25 µm, which corresponds to the electron spot size at a position 1 mm behind the jet exit, as expected from a few-µm size inside the gas[12] and the measured electron divergence. The spatial and temporal overlap of the driving and colliding lasers is ensured by observing the plasma self-emission from the collision zone from the top and time-resolved shadowgraphy (see methods) of the plasma from the side, the latter using a short probe pulse.

The X-ray beam is emitted from a source size that reflects the size of the electron bunch at the collision point and is confined in a cone given by the convolution of the electron beam divergence and a $1/(\gamma N_0^{1/2})$[24,25] radius cone due to relativistic beaming for small $a_0$, where $N_0$ denotes the number of laser oscillations, which in our case is approx. 10.

Figure 2 shows the X-rays obtained by 30MeV, 50MeV and 70 MeV electrons, leading to the emission of 15keV, 42keV and 83keV photons, respectively. A clear on-off behavior is evident, indicating that the X-rays indeed come from the electron-laser interaction and not from betatron oscillations during the acceleration process or Bremsstrahlung of electrons hitting the chamber walls. Due to their high energy, the photons were detected by an MCP-intensified scintillator fiber-coupled to a CCD-camera, which is not energy-selective, and hard to absolutely calibrate (see methods).

In order to obtain single-shot X-ray spectra, in the following a back-illuminated X-ray CCD camera operating in the single-photon-counting mode (see methods) replaced the scintillator camera. Since its sensitivity becomes impractically low beyond 35-40 keV, X-ray spectra are only presented for lower photon energies. The measured electron and corresponding X-ray spectra are shown in Fig.3. Normalized, run-averaged electron spectra are plotted as red lines in the left panel and X-ray spectra calculated from them as white lines in the right plot. Normalized, measured run-averaged photon spectra are shown as red lines in the right panel of Fig. 3 for comparison. Their shape seems to be in very good agreement with the expectation (see white lineouts in Fig.3 right) for a collision angle of 3.7° as used in the experiment. This angle was chosen to prevent the counter-propagating beam from shooting back into the laser system.

The main reason for shot-to-shot variations in photon numbers is the pointing instability of the laser, which due to the difference in focal length of the driver and colliding beams does not cancel out for both beams and leads to a varying beam overlap.

From Fig. 3 it is obvious that the photon energy scales approximately quadratic with the electron energy as expected from Equation (1).

This observation is elaborated in Fig.4, where the positions of the electron and X-ray spectral peaks and their corresponding rms spectral widths (in gray) are plotted for each shot in comparison with the expectation for different peak electric fields.

The data is in good agreement with the colliding pulse amplitude of $a_0$=0.9 inferred from the focal spot, laser duration and energy measurements, valid for a perfect temporal and spatial overlap between electrons and colliding beam. According to Equation 1, the X-ray spectral widths (vertical gray lines) may be attributed not only to the energy bandwidth of the electron bunch, but also to the temporarily varying intensity profile of the collision pulse which is Gaussian instead of top-head.

The inset in the figure shows the total electron numbers and x-ray photon-numbers/msr. Since the radial variation of the photon flux over the small CCD chip is negligible, it is not possible to extract X-ray divergence figures from the measurement for computing angle-integrated photon numbers.

Taking into account an estimated electron bunch duration of 5fs, as measured in similar experimental conditions[19,20] and our short colliding pulse, Eqn 3. yields an X-ray pulse duration of 5fs. We estimate the upper limit for the brilliance of the X-Ray source based on the measured X-Ray photon numbers and electron bunch emittance. Since the electrons come from a 2 μm source[12] with the directly measured divergence of 20...12 mrad FWHM one arrives at the values of $0.2\ldots5 \times 10^{20} \frac{photons}{s\,mm^2\,mrad^2\,0.1\%\,bandwidth}$, rising with electron energy, for the case of the interaction at the electron beam waist. In the experiment however the collision point was shifted 1.4mm downstream into the vacuum to ensure the unambiguous collision measurements. At this position the electron bunch transversely expands to 20um, yielding the reduced values of $0.2\ldots5 \times 10^{18} \frac{photons}{s\,mm^2\,mrad^2\,0.1\%\,bandwidth}$. It is worth mentioning that the interaction position is easily tuned by delaying the collision pulse, so the former brilliance estimates can fairly be attributed to the source demonstrated.

With better detection systems suitable for higher energy photons, the usable energy range of our source can be readily expanded. Our shock-front acceleration scheme routinely delivers quasi-monoenergetic electron bunches with energies tunable in the range of 15-150MeV[16], which would cover an accessible photon energy range from 5keV-500keV with the present setup.

Further developments will center on optimizing the collision parameters. The colliding pulse can be carefully tailored[27] to achieve the small collision radius, high colliding flux, but low intensity in order to enhance the scattered X-ray flux while keeping its spectrum narrow.

By combining this concept with future high-repetition rate laser systems based on OPCPA amplification and high-repetiton rate thin-disk pump lasers[28,29], such a Thomson source could serve a variety of imaging applications in research, medicine and industry due to its compactness, tunability and unmatched short pulse duration.

**Methods**

Laser-wakefield Acceleration of Electrons

LWFA is well known and documented in many original and several review articles (e.g. [30]), so we will only give a short description of the injection concept that controls the electron energy here. Moreover, the regime of electron acceleration used for this work is covered in detail in [16].

When an intense laser pulse ($a_0>1$) travels through an underdense plasma, it drives a plasma wave by pushing the electrons aside with its ponderomotive force. The phase velocity of this wave equals the laser´s group velocity, which is close to speed of light. The wave therefore constitutes a fast-moving longitudinal field structure that accelerates electrons that get trapped in it. Self-trapping can occur e.g. by nonlinear wave-breaking, which due to the stochastic nature of the injection phase commonly results in a large energy spread and poorer reproducibility. Therefore, self-trapping was avoided in favor of a forced trapping scheme called shock-front injection [16, 26]. Placing the edge of a razor blade into the supersonic flow of a de Laval nozzle creates a shock-front. This causes a sharp density drop of the order of the molecular mean free path ($1.3 \mu m$) in laser propagation direction. When the plasma wave crosses this drop, its wavelength instantaneously increases. This places the former first wave crest into the trough of the now elongated wave, into a position with a strong accelerating field. Therefore a large number of electrons from the crest are trapped at a defined position and undergo the same acceleration in the downstream part of the wave, all gaining similar energy. The position of the shock in a limited-length gas target determines the remaining acceleration length and hence the final energy. If the plasma density is kept safely below the self-injection limit (experimentally determined by lowering the density such that without the razor blade no electrons are observed), no further electrons are being injected and the absolute energy spread of the beam is kept low.

In our experiment we used a 300μm-diameter supersonic nozzle exhausting helium gas into the chamber vacuum. With a backing pressure of 14 bar it produces an electron density of $5 \times 10^{18} cm^{-3}$ at the position of the laser. At this density spontaneous wave breaking and self-injection can be avoided even at a peak laser amplitude of $a_0$=4.4, which resulted from focusing the 28 fs, 1.2 J laser pulses to a FWHM diameter of 13 µm using an f/13 off-axis parabola. The electron beam was analyzed in a 1 T permanent magnet setup described in [31].

Colliding beam setup

The collision position was chosen to be 1mm after the nozzle exit. This distance was long enough to keep the colliding beam from disturbing the electron acceleration and to keep the interaction point in near-vacuum, but short enough to avoid a large transverse electron bunch spread due to its divergence.

With a pulse energy of 300mJ, a spot size of 25 µm and a 28-fs pulse duration, we deduce a peak $a_0$ of 0.9, constraining Thomson scattering to the mostly linear regime[16].

Shadowgraphy was used to ensure the temporal alignment of the colliding pulse with the electron bunch. A small part of the laser beam, timed by a delay stage, acts as a transverse probe of the interaction. The electron density modulations caused by the propagation of drive and colliding pulse leave an imprint in the probe wavefront and lead to intensity modulations reflecting the electron density gradients. Scanning the delay of the probe beam allows to track the progression of the ionization fronts. By observing both drive and colliding beam separately, the collision point relative to the electron injection and acceleration regions can be precisely determined.

For transverse alignment of both beams a top- and a side-view imaging system was used that monitor the Thomson side-scattered light from the background plasma electrons which allowed us to monitor and overlap the beams at the supposed collision position.

X-ray detection

We used two different X-ray cameras to characterize the source. One at a time, they were installed on the laser propagation axis, 2.4 meters after the nozzle (see fig 1). Electrons were deflected by the spectrometer, and a 30µm aluminum foil blocked residual laser- and stray-light.

For Fig.2, a Proxitronic scintillator-based intensified X-ray camera was used. It couples a 2" phosphor screen to a 1/2" Allied Marlin CCD camera via two optical fiber reducer stages and a Chevron-type MCP image intensifier with variable gain. It offers spatial resolution and high sensitivity within a broad spectral range between 2.5keV and 100keV, but no absolute information about the photon energy due to variable gain.

For Figures 3 and 4, Single-Photon-Counting-Spectroscopy (SPCS) using an Andor DO432 BN-DD back-illuminated X-ray CCD camera with 1250x1152 22.5µm-size pixels yielded single-shot X-ray spectra. This technique exploits the fact that camera readout counts are directly proportional to the absorbed X-ray photon energy in each pixel. In practice, the energy of a single photon is deposited within several (up to 3-4) neighboring pixels ('clusters'). In a post-processing step the content of all these pixels is added up to yield the photon energy, provided the clusters are well separated laterally. This sets a limit on the number of photons detected in a single shot, which in our case, according to Poisson statistics, is 40000 if the probability of a cluster being caused by two photons is kept below 0.5%. The source-detector distance was chosen such that this condition is fulfilled. The histograms of the post-processed images corrected with the camera sensitivity and filter transmission curves (for 30µm Aluminum, 250µm Kapton and 3cm air) yield the absolute X-ray spectrum of the source, averaged over the observed solid angle (in our case 0.13 msr corresponding to a 28.1x25.9mm$^2$ chip in a 2.4m distance). As the quantum efficiency of the direct detection almost vanishes above ~40keV, only the spectra of the runs delivering X-rays below this energy are presented. Photon numbers are given per solid angle, since the full beam divergence is considerably larger than the CCD chip for lower energies, making an estimate of total photon yield difficult.

## Acknowledgment

This work was supported by DFG through the MAP and TR-18 funding schemes, by EURATOM-IPP, and the Max-Planck-Society.

## Author contribution

K.K., A.B., L.V. and S.K. designed the experiment. K.K., J.W., J.X. and A.B. carried out the measurements. K.K. and S.K. wrote the main part of the paper. L.V. and S.K. provided overall guidance and supervised the project. All authors discussed the results, reviewed and commented on the manuscript.

# Author information

Correspondence and request for materials should be addressed to S.K. (stefan.karsch@mpq.mpg.de).

# Figure Captions:

Figure 1. Experimental setup. 0.8µm-wavelength, 28fs-duration laser pulses enter the target chamber and are split by a pickoff mirror into a driver (1.2J) and colliding beam (0.3 J). The driver is focused to an intensity of $4.2 \times 10^{19}$ W/cm$^2$ into a supersonic helium gas jet from a 300µm-diameter nozzle equipped with a razor blade to create a shock front, where it accelerates quasi-monoenergetic electron beams via shock-injected LWFA. These are analyzed in a calibrated 1T-dipole magnet spectrometer with a scintillating screen observed by a camera. The colliding beam is focused to a point 1 mm behind the gas jet to reach an intensity of $1.75 \times 10^{18}$ W/cm$^2$, corresponding to $a_0=0.9$ and a therefore mostly linear interaction. It collides at an angle of 3.7° with the electron beam and acts as an optical undulator, leading to emission of X-rays. They are detected after passing through a 30-µm aluminum light-block filter and 250 µm Kapton vacuum window by either a scintillator-based or single-photon counting X-ray CCD camera mounted on the electron propagation axis as defined before the deflection magnet.

Figure 2. Images of the X-ray beam obtained with a scintillator-based MCP-intensified camera. Series a, b and c show 4 typical intensified images (top) from the runs with different electron energies of 30,50 and 70MeV respectively. According to eq (1), these electron energies correspond to X-ray beam peak photon energies of 15, 42 and 83keV respectively. 2 lower images in each section are the reference shots obtained by blocking only the counter-propagating beam. The gain of the MCP was doubled for the series c compared to a and b in order to compensate for the reduction

of the phosphor scintillator sensitivity at this high energy. Due to enhanced beaming, the brightness of series a and b seems to be equal in spite of an 8-fold reduction in the scintillator sensitivity at the higher energy in b.

Figure 3. Direct measurements of single-shot electron (left) and corresponding X-ray photon spectra (right). Every single horizontal trace corresponds to a single laser shot. Shown are the best 50% of shots by photon number in each run. Different vertical sections represent experimental runs with different razor-blade positions and hence different electron beam energy settings. Color-coded on the right plot are the absolute X-ray photon spectra directly measured by a 28.1x25.9mm$^2$ camera chip at a 2.4m distance from the source corresponding to the collection of radiation emitted within 0.13msr solid angle. The photon spectra are corrected for the transmission of the laser blocking filter, the vacuum window (see Fig. 1) and the CCD sensitivity curve. Normalized run-averaged spectra are shown in overlaid red lines. White lines in the right picture show the X-ray spectra theoretically expected for the corresponding mean electron spectra left. The simulation was performed with the SPECTRA 9.0[32] code.

Figure 4. Spectral peaks for all single shots, sorted by electron energy settings. The peak position was determined by fitting a Gaussian function to the measured spectra, whose rms-widths define the lengths of the error bars. The colors distinguish the runs with different shock positions and indicate the reproducibility and resolution of each setting. Two parabolic curves (blue solid lines) show the expected X-ray peak positions according to equation 1 for a peak $a_0$ of 0.83 and 0.0 in the colliding beam, the former being the best fit of $a_0$ in equation 1 to the measured X-Ray spectra. The inset shows the measured electron and X-ray photon/msr numbers for each shot.

**Figures:**

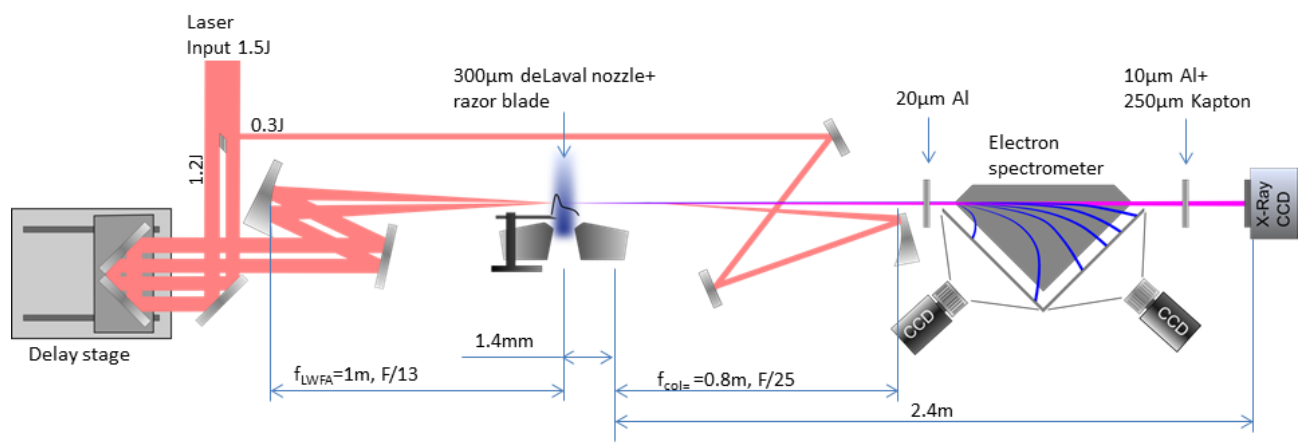

Fig 1.

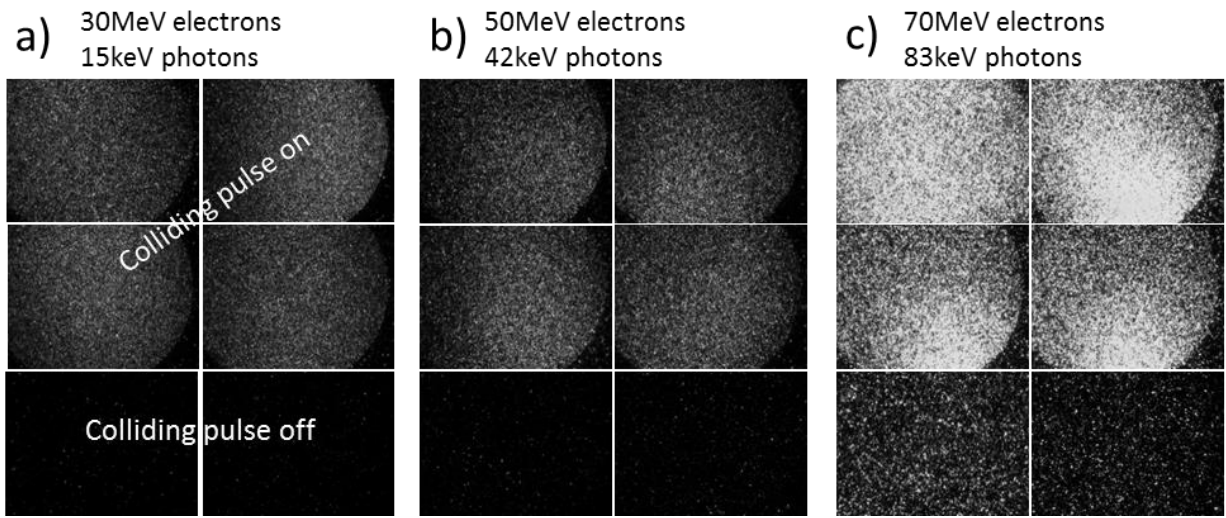

Fig 2.

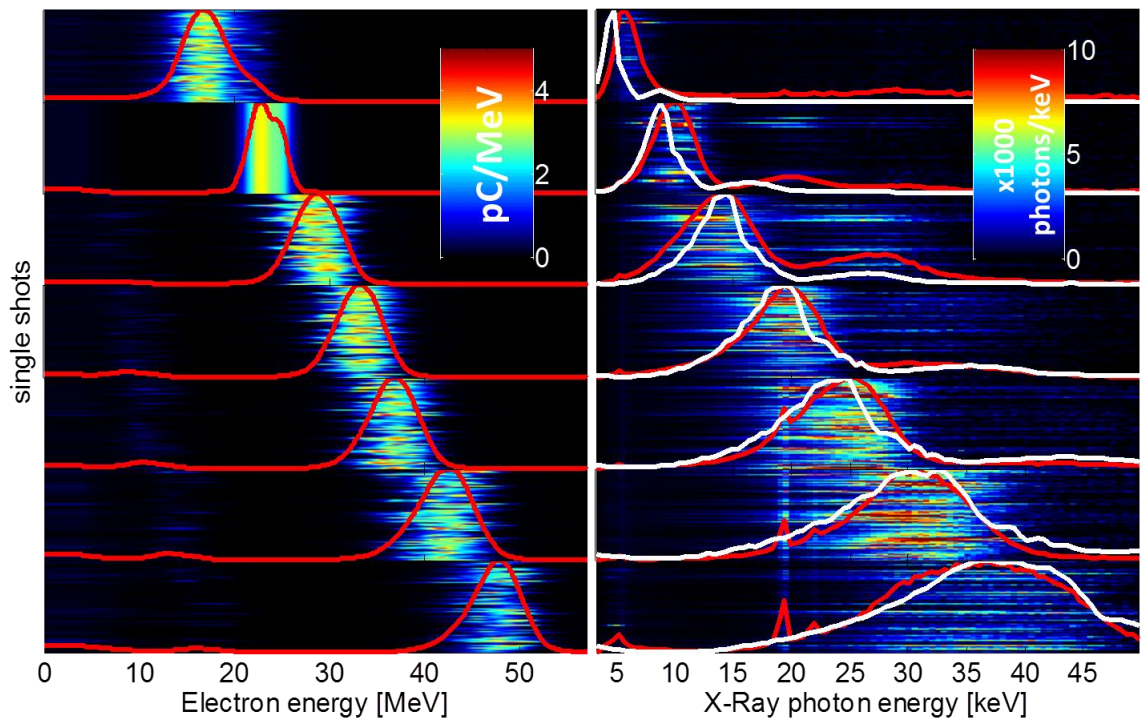

Fig. 3

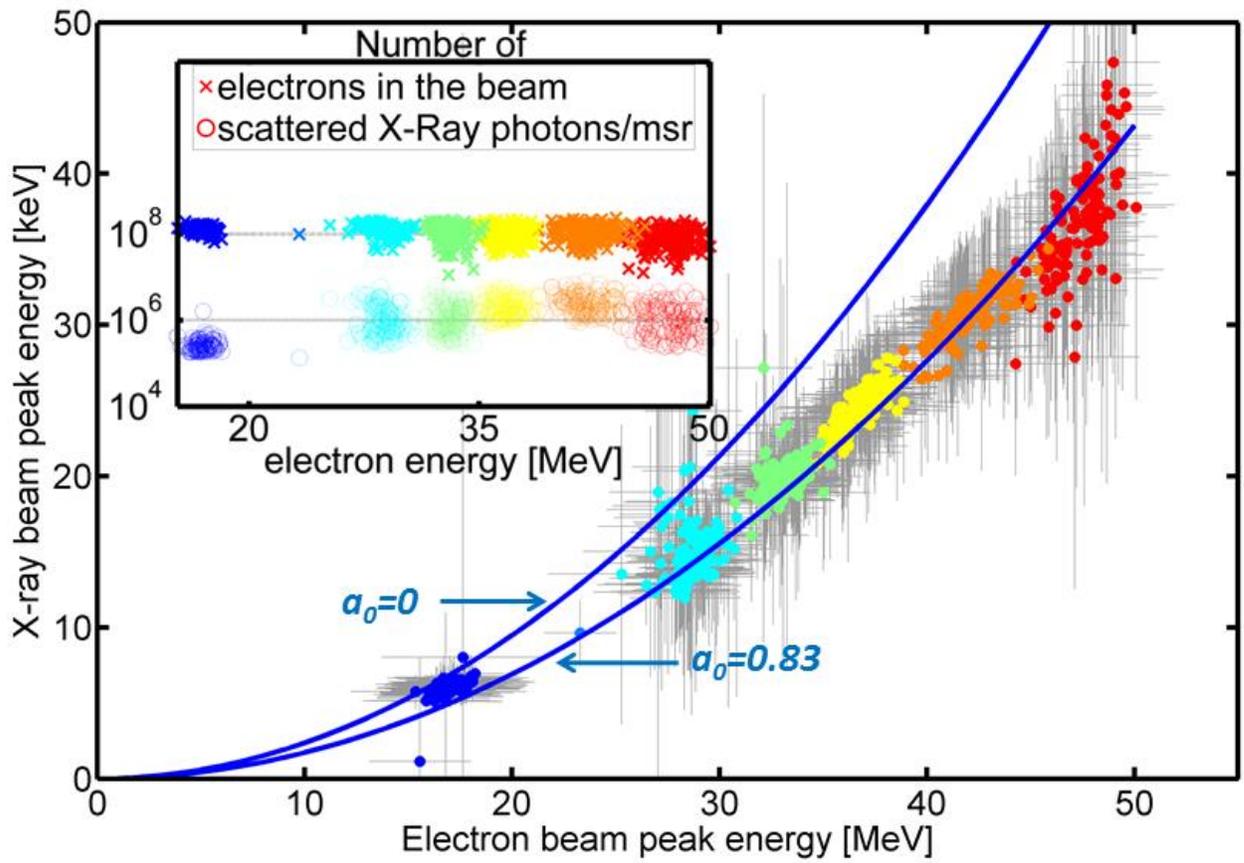

Fig. 4